\documentclass[preprint,tightenlines,eqsecnum,floats,aps,amsmath,amssymb,nofootinbib,prd,showpacs,superscriptaddress]{revtex4}  
\usepackage{amsmath,amssymb,amsfonts}
\usepackage{graphicx}
\usepackage{enumerate}
\usepackage[colorlinks=true,linkcolor=blue,anchorcolor=blue,citecolor=blue,urlcolor=blue]{hyperref}
\usepackage[sort&compress]{natbib}
\usepackage{url}
\usepackage[T1]{fontenc}
\usepackage{mathptmx}
\usepackage{float}
\usepackage{booktabs}
\usepackage{CJK}
\usepackage{epstopdf}
\usepackage{tikz}
\usetikzlibrary{positioning, shapes.geometric, arrows}

\begin{document}

\title{ Precise constraint on properties of neutron stars through new universal relations and astronomical observations}

\author{Zehan Wu }
\affiliation{School of Physics and Optoelectronics, South China University of
Technology, Guangzhou 510641, P.R. China}
\author{ Dehua Wen\footnote{Corresponding author. wendehua@scut.edu.cn}}
\affiliation{School of Physics and Optoelectronics, South China University of
Technology, Guangzhou 510641, P.R. China}
\date{\today}

\begin{abstract}
In view of the great uncertainty of the equation of state (EOS) of high-density  nuclear matter, establishing EOS-independent universal relations between global properties of neutron stars provides a practical way to constrain the unobservable or difficult-to-observe properties through astronomical observations. It is common to construct universal relations between EOS-dependent properties (e.g., moment of inertia, tidal deformation, etc.) or combined properties (e.g., compactness). Improving the precision of the universal relations may provide stricter constraint on the properties of neutron star.  We find that in 3-dimensional space with mass and radius as the base coordinates,  the points corresponding to a certain property of neutron star described by different EOSs are almost located in the same surface. Thus the universal relation between the property and the stellar mass-radius can be expressed through describing the surface.
 It is shown that the resulting universal relations have higher precisions.  As an example, we construct high-precision universal relations for the moment of inertia, the $f$-mode frequency, and the dimensionless tidal deformation respect to the mass-radius. As the observational data of  neutron star mass and radius from NICER  grows in data and accuracy, these  universal relations allow for more precise constraints on the unobservable or difficult-to-observe properties.

\end{abstract}


\maketitle

\section{Introduction}
Neutron stars are one of the densest objects in the universe. The structure and properties of neutron stars are sensitive to EOS, making them ideal laboratories for testing theories of dense matter physics \cite{Breu_MNRAS2016, Yunes_NatRP2022, Lattimer_ARONPS2021, Huth_Nat2022}. However, due to the uncertainty of the EOS, many neutron star unobservable or difficult-to-observe properties are hard to predict precisely. Fortunately, groundbreaking observations from detectors such as NICER, LIGO, and VIRGO have provided new insights into the properties of neutron stars. Since the observation of GW170817 \cite{Abbott_PRL2018, Abbott_APJL2017}, there have been many promising results from astronomical observations of neutron stars, such as the PSR J0030+0451 \cite{Miller_APJL2021}, PSR J0740+6620 \cite{Riley_APJL2021}, GW190814 \cite{Abbott_APJL2020}, etc., which provided the possibility of constraining unobservable properties of neutron stars and even further constraining the EOS of high-density asymmetric nuclear matter \cite{Richter_PRC2023, Baiotti_PPNP2019}.

There are several ways to achieve constraints on the properties of neutron stars based on astronomical observations. Bayesian analysis methods were one of the practise way. For example, through this method, Steiner $et ~al.$  inferred the radius range of a canonical neutron star from six neutron star observations \cite{Steiner_APJ2010}; G\"uven $et ~al.$ deduced both the EOS and the canonical neutron star radius \cite{Guven_PRC2020}. Universal relations provide another way to constrain the properties of neutron star through  astronomical observations. This method establish  EOS-independence relations between the global properties of neutron stars.
Early in 1994, Ravenhall and Pethick proposed the universal relation between the normalized moment of inertia $I/MR^2$ and the compactness $C$ \cite{Ravenhall_APJ1994}. This relation was refined by Lattimer \& Prakash \cite{Lattimer_APJ2001} and Bejger \& Haensel \cite{Bejger_AA2002}, and later used by Lattimer \& Schutz to estimate the radius of neutron stars in binary systems \cite{Lattimer_APJ2005}. Breu and Rezzolla switched to the dimensionless moment of inertia $I/M^3$ to establish a more precise relation with $C$ \cite{Breu_MNRAS2016}, which was confirmed at the maximum rotation sequence by Ref. \cite{LJJ_PRC2023}.  In 1998, Andersson and Kokkotas proposed a classical universal relation between the frequency of the $f$-mode $\omega_f$ and the average density of a neutron star in the field of asteroseismology \cite{Andersson_MNRAS1998}. This relation is still received much attention in recent years \cite{Benhar_PRD2004, Doneva_PRD2013, Chirenti_PRD2015, Wen_PRC2019, Pradhan_PRC2021}. In 2013, notable relations $I\ -\ LOVE\ -\ Q$ established by Yagi and Yunes \cite{Yagi_Science2013}. Pappas and Apostolatos found that this relation breaks down under rapidly rotating conditions \cite{Pappas_PRD2014}. Soon after, Doneva $et ~al.$ and Chakrabarti $et ~al.$  proposed the $I\ -\ LOVE\ -\ Q$ relation for rotating neutron stars and rotating quark stars \cite{Doneva_APJ2013, Chakrabarti_PRL2014}, respectively. In more recent years, the relation between the macroscopic properties of neutron stars and the nuclear matter parameters was established \cite{Yang_PRD2023, Richter_PRC2023}, and  the universal relations related to the rotational parameters of neutron stars are also introduced \cite{Cipolletta_PRD2017, Luk_APJ2018, Papigkiotis_PRD2023, Kruger_PRD2023} .

With the help of the notable $I\ -\ LOVE\ -\ Q$ relations, the measurement of the moment of inertia of PSR J0737-3039A can obtain the radius and tidal deformation with precise mass data, which is of great significance for the dense matter EOS \cite{Lattimer_ARONPS2021}. However, Sulieman and Read emphasize that these existing universal relations still suffer from insufficient precision in constraining the properties of neutron stars, especially the radius \cite{Suleiman_PRD2024}. Is it possible to establish a more precise universal relation?

As we know, the properties of neutron stars exhibit EOS dependence \cite{Bejger_AA2002, Papigkiotis_PRD2023}, and there is a one-to-one correspondence between the global properties mass-radius of neutron star and the EOS pressure-density of dense nuclear matter  \cite{Lindblom_APJ1992}. In a sense, the EOS dependence is equivalent to mass-radius dependence.  For different EOSs, we find that the points corresponding to a certain property of neutron star  are almost located in the same surface in 3-dimensional space with mass and radius as the base coordinates. Through describing the surface, we can get the  universal relation between the property and mass-radius.
 It is shown that the resulting universal relations have higher precisions. We called this method as mass-radius projection.
In this work, through the mass-radius projection method, we  establish several  high-precision universal relations which related to the moment of inertia, the $f$-mode frequency and the  tidal deformation. Based on the observation data of neutron star mass and radius from NICER, the related  properties of the observed neutron stars   can be constrained more precisely.

This paper is organized as follows. Sec. \ref{s1} introduces the EOS model and the basic formulas for neutron star properties. In Sec. \ref{s2},  the universal relations of static-neutron-star properties  are established through the mass-radius projection method, where Sec. \ref{s2a}, Sec. \ref{s2b}, and Sec. \ref{s2d} present the   universal relation of the moment of inertia, the $f$-mode frequency, and  tidal deformation, respectively. Particularly, Sec. \ref{s2c} introduces the application of affine transformation in the establishment of universal relations, where universal relation between $I$ and $\omega_f$ established as an example. The summary of this work drawn in Sec. \ref{s3}.

Unless otherwise stated, in the formula we use the gravitational units which $G = c = 1$.

\section{Equation of State and neutron star properties} \label{s1}

\subsection{Equation of state model}

A set of representative EOSs is necessary to prove the existence of a universal relation. This includes: (1) the piecewise-polytropic (PP) model \cite{Muller_AA1985}; (2) the microscopic and phenomenological nuclear many-body model, such as APR \cite{Akmal_PRC1997}, APR3 \cite{Akmal_PRC1997}, APR4 \cite{Akmal_PRC1997}, ALF2 \cite{Alford_APJ2005}, ENG \cite{Engvik_apj1996}, DDLZ1 \cite{Wei_cpc2020}, DD-ME1 \cite{Nik_PRC2002}, DD-ME2 \cite{Lalazissis_PRC2005}, MPA1 \cite{Muther_plb1987}, PKA1 \cite{Long_prc2007}, PKO3 \cite{Long_prc2007}, SLy \cite{Douchin_aa2001}, WFF1 \cite{Wiringa_prc1988}, WFF2 \cite{Wiringa_prc1988}, soft-EOS \cite{Hebeler_apj2013}, and stiff-EOS \cite{Hebeler_apj2013}; (3) the first-order phase transitions model, which combines APR3 with quark EOS and DD-ME2 with quark EOS, exhibiting first-order phase transitions at densities of $1\ \rho_0$ (saturation density), $1.5\ \rho_0$, $2\ \rho_0$.

\begin{figure}[H]
    \centering
    \includegraphics[scale=0.4]{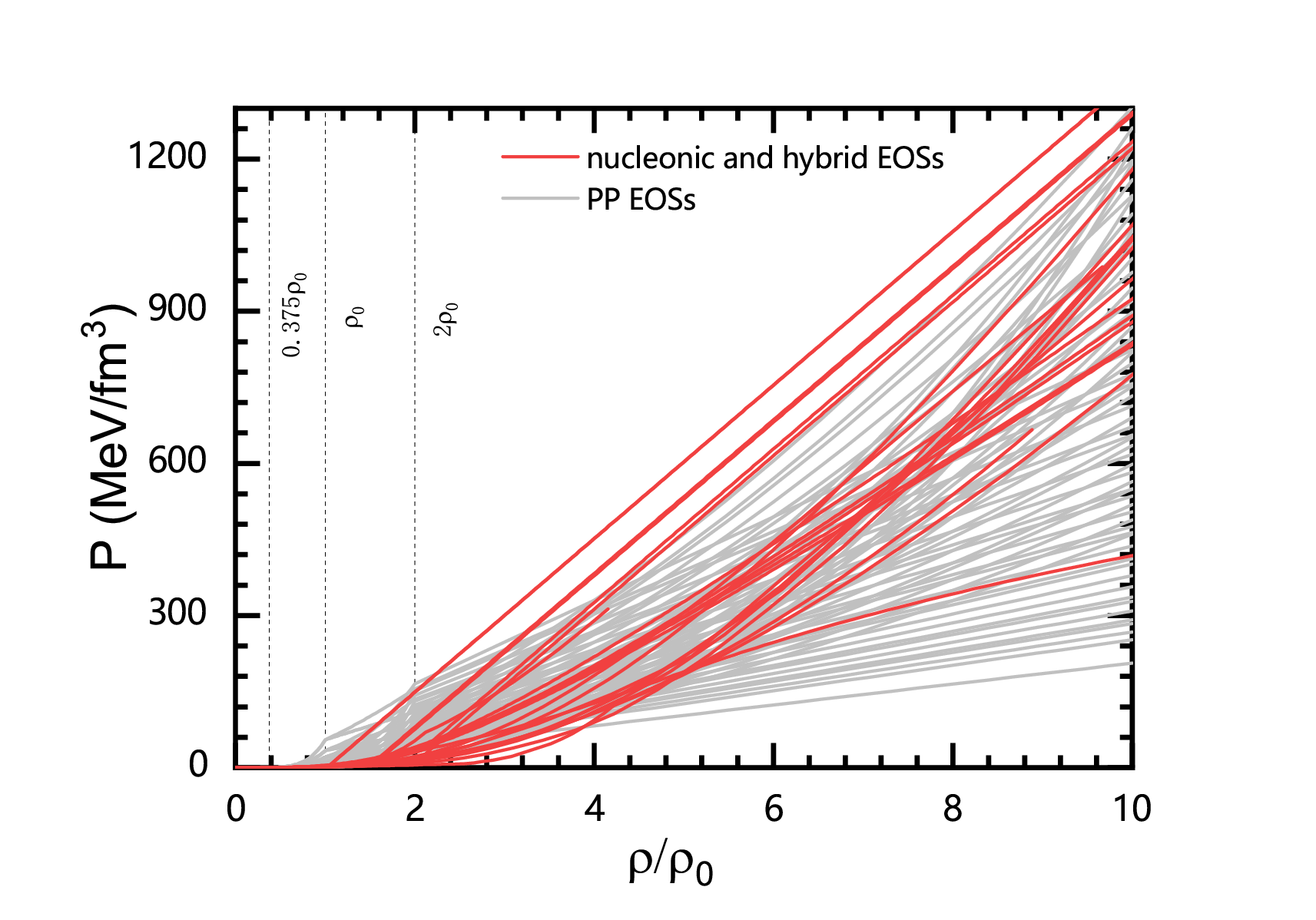}
    \caption{The $P-\rho$ relations of the adopted EOS models, where $\rho_0$ denotes the saturation density. The gray curves represent PP EOSs, while the vertical dashed lines indicate the transition of points between different PP segments. The red coloured curves represent nucleonic and hybrid EOSs. }
    \label{1}
\end{figure}

For the PP model, each segment is represented by a different density $\rho _i$ and pressure $P_i$, which can be written as
\begin{equation}
P=K_i\rho ^{\varGamma _i}\,\,\left( \rho _i\leqslant \rho \leqslant \rho _{i+1} \right),\label{1.1}
\end{equation}
where $K_i$ is the polytropic constant defined as
\begin{equation}
K_i=\frac{P_i}{\rho _{i}^{\varGamma _i}},\label{1.2}
\end{equation}
and the adiabatic index $\varGamma _i$ defined as
\begin{equation}
\varGamma _i=\frac{\log _{10}\left( P_{i+1}/P_i \right)}{\log _{10}\left( \rho _{i+1}/\rho _i \right)}.\label{1.3}
\end{equation}

Here, we use BPS+NV EOSs \cite{Baym_APJ1971, Negele_NPA1973} as the crust segment. The core segment is parameterized starting at 0.375 $\rho_0$ with an adiabatic index $\varGamma _i$ in the range $\left[ 1, \ 5 \right]$. Figure \ref{1} shows the EOSs included in our collection, which can be seen as broadly representative from the distribution of pressure-density. By the way, the causality condition ($\sqrt{dP/d\rho}\leqslant 1$) and maximum mass condition ($M_{max}\geqslant 2.14\ M_{\odot}$ \cite{Cromartie_NatA2020}) are satisfied for the constructed 63 piecewise-polytropic EOSs.

\subsection{Basic formulas for neutron star properties}
The moment of inertia of the neutron star reflects the particles content of matter, the mass associated with the energy required to compress the matter to a given density, and the effective energy of the gravitational interactions between different parts of the star \cite{Hartle_APJ1968}. In General Relativity, the moment of inertia is defined as
\begin{equation}
I=J/\varOmega, \label{2.1}
\end{equation}
where $J$ and $\varOmega$ represent the angular momentum and the spin frequency, respectively. We adopt the moment of inertia of spherically symmetric neutron stars in general relativity from Ref. \cite{Dong_PRD2023} in calculations, expressed as

\begin{equation}
\frac{dI}{dr}=\frac{8}{3}\pi r^4\rho \left( 1+\frac{P}{\rho} \right) \left( 1-\frac{5I}{2r^3}+\frac{I^2}{r^6} \right) \left( 1-\frac{2m}{r} \right) ^{-1}, \label{2.2}
\end{equation}
where $m$ is the mass as a function of radius $r$, given by the TOV equation. This approach has been checked to be in basic agreement with the results of the conventional algorithm \cite{Hartle_APJ1968, Hartle_APJ1967}.

The tidal deformation $\lambda$ is defined by the star's induced quadrupole moment $Q_{ij}$ in response to the external tidal field $\varepsilon_{ij} $,
\begin{equation}
Q_{ij}=-\lambda \varepsilon _{ij}. \label{b.1}
\end{equation}
The second tidal-Love-number $k_2$ is defined as
\begin{equation}
k_2=\frac{3}{2}\lambda R^{-5}, \label{b.2}
\end{equation}
which can be solved together with the TOV equation \cite{Hinderer_APJ2008, Hinderer_PRD2010}. Once $k_2$ is determined, the dimensionless tidal deformation $\varLambda$ can be determined by the relation
\begin{equation}
\varLambda =\frac{2k_2}{3C^5}, \label{b.3}
\end{equation}
where $C=M/R$ is the compactness.

The $f$-mode  is the lowest-frequency quadrupole oscillation  of neutron stars \cite{Lindblom_APJ1983}. The basic formulas to calculate $f$-mode oscillation can be found in Refs. \cite{Regge_PR1957, Lindblom_APJ1983, Detweiler_APJ1985}. The numerical solution method used in this work is the Detweiler-Lindblom method \cite{Lindblom_APJ1983, Detweiler_APJ1985}.

\section{Establishing the universal relations through  mass-radius projection} \label{s2}
 It is well known that there is a one-to-one correspondence between the properties mass-radius of neutron stars and the EOS pressure-density of dense nuclear matter \cite{Lindblom_APJ1992}. This means that the EOS-dependent neutron star properties also embodies mass-radius dependence.
   We find that for different EOSs the points corresponding to a certain property of neutron star  are almost located in the same surface in 3-dimensional space with mass and radius as the base coordinates. Through describing the surface, we can get the  universal relation between the property and mass-radius. Our calculation shows that the resulting universal relations have higher precisions.
  In this section, through the mass-radius projection of the moment of inertia, the frequencies of $f$-mode, and the tidal deformation, the precise universal relations of these properties with the mass-radius of neutron stars are presented.  Based on these universal relations, combining with the astronomical observations from NICER, such as the mass and radius constraints of PSR J0030+0451 \cite{Miller_APJL2021} and PSR J0740+6620 \cite{Riley_APJL2021}, a relative precise prediction of the related properties of these observed neutron stars are given.

\subsection{The universal relation for the moment of inertia }\label{s2a}

Many universal relations have been constructed for the moment of inertia, such as the well-known $I\ -\ LOVE\ -\ Q$ \cite{Yagi_Science2013, Doneva_APJ2013, Chakrabarti_PRL2014}, $I\ -\ C$ \cite{Ravenhall_APJ1994, Lattimer_APJ2001}, $I\ -\ BE$ \cite{Steiner_EPJA2016}, where $BE$ is the binding energy. It is always necessary to take transformed quantities such as $I/M^3$ and $I/R^3$ for $I$ in these universal relations. One example is that Breu and Rezzolla improved the precision of predicting the moment of inertia by using $I/M^3$ in the universal relation \cite{Breu_MNRAS2016}. Her  we try to build a more precise universal relation related to the moment of inertia through the mass-radius projection method. In addition, considering that the mass and radius of neutron star are observable,  it is possible to provide more precise constraints on the moment of inertia of the observed neutron stars.

\begin{figure}[H]
    \centering
    \includegraphics[scale=0.45]{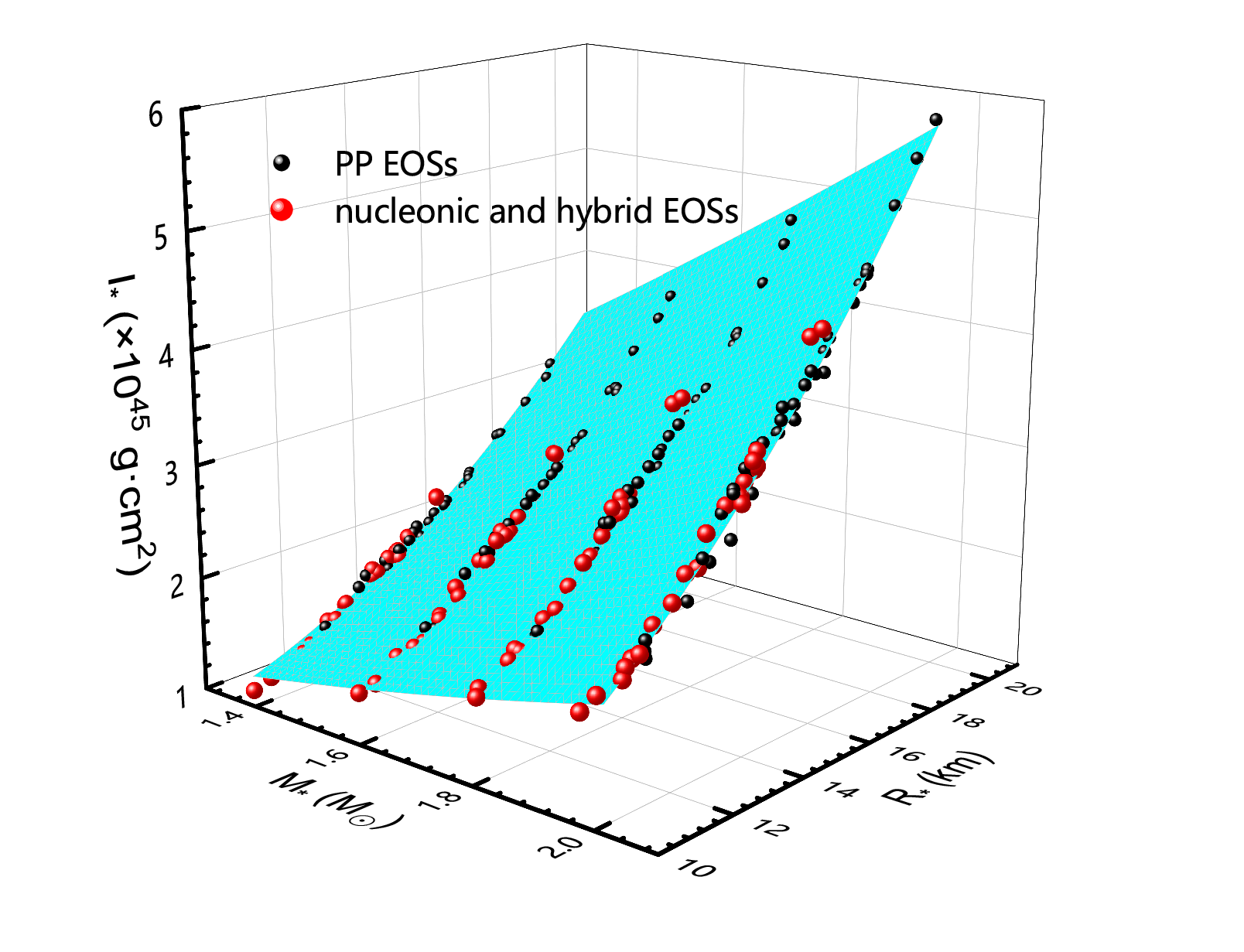}
    \caption{The moment of inertia as a function of the stellar mass and radius, where the black dots represent the results of the piecewise-polytropic EOSs, the red dots represent the results of nucleonic and hybrid EOSs, and the cyan surface is the fitting surface.}
    \label{2}
\end{figure}

Figure \ref{2} illustrates the values of the moment of inertia (globular dots) as a function of the the mass and radius in 3-dimensional space. It is shown that for different EOSs, almost all of the points are located in the same fitting surface. These means that the moment of inertia can be  described accurately by the two independent variables, mass and radius.  Through describing the surface, we can get the  universal relation between the moment of inertia and mass-radius as

\begin{equation}
I_*=4.417-2.696M_*-0.513R_*+0.427{M_*}^2+0.014{R_*}^2+0.249M_*R_*, \label{3.1}
\end{equation}
in term of dimensionless parameters  $I_*=\frac{I}{10^{45}\ g\cdot cm^2}$, $M_*=\frac{M}{M_{\odot}}$, and $R_*=\frac{R}{\mathrm{km}}$.

\begin{figure}[H]
    \centering
    \includegraphics[scale=0.45]{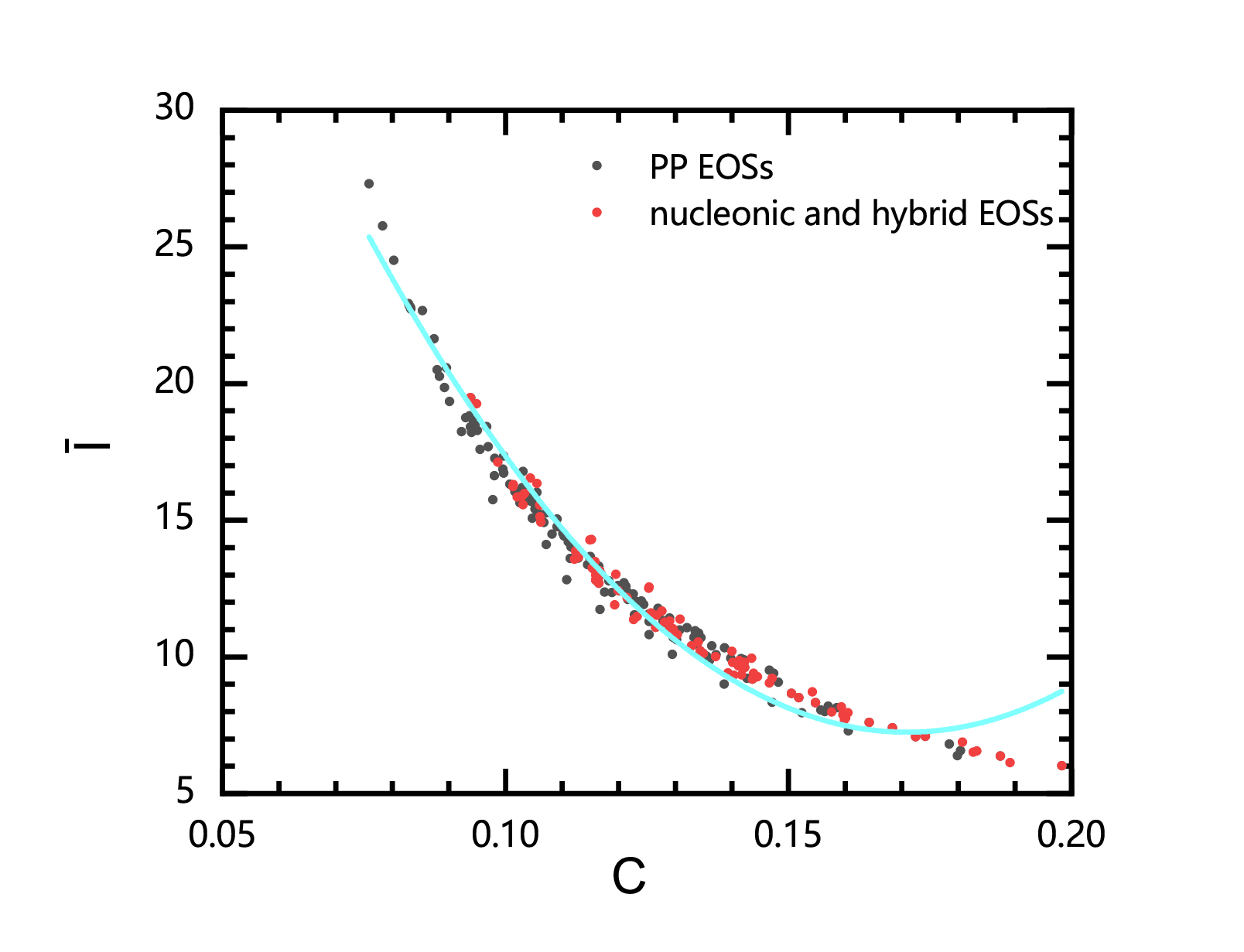}
    \caption{$\overline{I}$ - $C$ relation with different EOSs together with the fitting curve (cyan). Black dots represent the results of the piecewise-polytropic EOSs. Red dots represent the results of nucleonic and hybrid EOSs.}
    \label{3}
\end{figure}

In order to examine the precision of $I_*\ -\ \left( M_*,\ R_* \right)$, we choose the commonly used universal relation between the dimensionless moment of inertia $\overline{I}(=I/M^3)$ and compactness $C$ for comparison \cite{Ravenhall_APJ1994, Lattimer_APJ2001}. Figure \ref{3} shows the $\overline{I}\ -\ C$ universal relation with the same EOSs set. By using a polynomial of the same order, the universal relation $\overline{I}\ -\ C$ can be expressed as
\begin{equation}
\bar{I}=65.857-685.567C+2004.727C^2. \label{3.2}
\end{equation}

To quantitatively compare the accuracy of these two universal relations, the means of the relative errors are presented in Table  \ref{t1}. It is clear that the universal relation described by Eq. \ref{3.1} has much better precision.

\begin{table}[H]
\caption{\textbf{Mean of relative errors for $I_*\ -\ \left( M_*,\ R_* \right)$ and $\overline{I}\ -\ C$, where $X_i$ represents the value of the relative error.}} \label{t1}
\centering
\begin{tabular}{lll}
\toprule
&~~~~~~~~~$I_*\ -\ \left( M_*,\ R_* \right)$&~~~~~~~~~$\overline{I}\ -\ C$ \\
\midrule
Mean $\left( \frac{1}{n}\sum_{i=1}^n{X_i} \right)$ &~~~~~~~~~0.019&~~~~~~~~~0.774 \\
\bottomrule
\end{tabular}
\end{table}

According to Eq. (\ref{3.1}) and the observed data of stellar mass and radius from NICER, a relative precise constraint on the moment of inertia for the observed neutron star can be predicted. For PSR J0030+0451 with  mass of $1.44\pm 0.15\ M_{\odot}$ and radius of $\left[ 11.2- 13.3 \right]\ \mathrm{km}$ at 90\% credibility  \cite{Miller_APJL2021}, its  moment of inertia can be constrained in a range of $\left[0.980-3.408\right] \times{10^{45}\ \textrm{g}\cdot \textrm{cm}^2}$. This result is in basic agreement with the prediction ($1.43_{-0.13}^{+0.3}\ \times{10^{45}\ \textrm{g}\cdot \textrm{cm}^2}$) in Ref. \cite{Jiang_APJ2020}. Similarly, for PSR J0740+6620 with mass of $2.08\pm 0.07\ M_{\odot}$ and radius of $\left[ 12.2-16.3 \right]\ \mathrm{km}$ at 95\% credibility \cite{Riley_APJL2021}, its  moment of inertia should be in a range of $\left[1.859- 7.364\right]\times{10^{45}\ \textrm{g}\cdot \textrm{cm}^2}$.

\subsection{The universal relation for the frequencies of $f$-mode} \label{s2b}

In recent years, many universal relations have been established for the frequencies of $f$-mode, such as $\omega_f\ -\ D _t$ \cite{Wen_PRC2019}, $\omega_f\ -\ C$ \cite{ZhaoTQ_PRD2022}, $\omega_f\ -\ \varLambda $ \cite{Sotani_PRD2021}, where $\omega_f$ and $D _t$ are the frequency and damping time of  $f$-mode, respectively. In particular, the universal relation $\omega _f\left( \mathrm{kHz} \right)=a+b\sqrt{M_*/{R_*}^3}$ has received long and continuous attention, where constant $a$ and $b$ give different results according to different studies \cite{Andersson_MNRAS1998, Benhar_PRD2004, Doneva_PRD2013, Chirenti_PRD2015, Wen_PRC2019, Pradhan_PRC2021}.

\begin{figure}[H]
    \centering
    \includegraphics[scale=0.5]{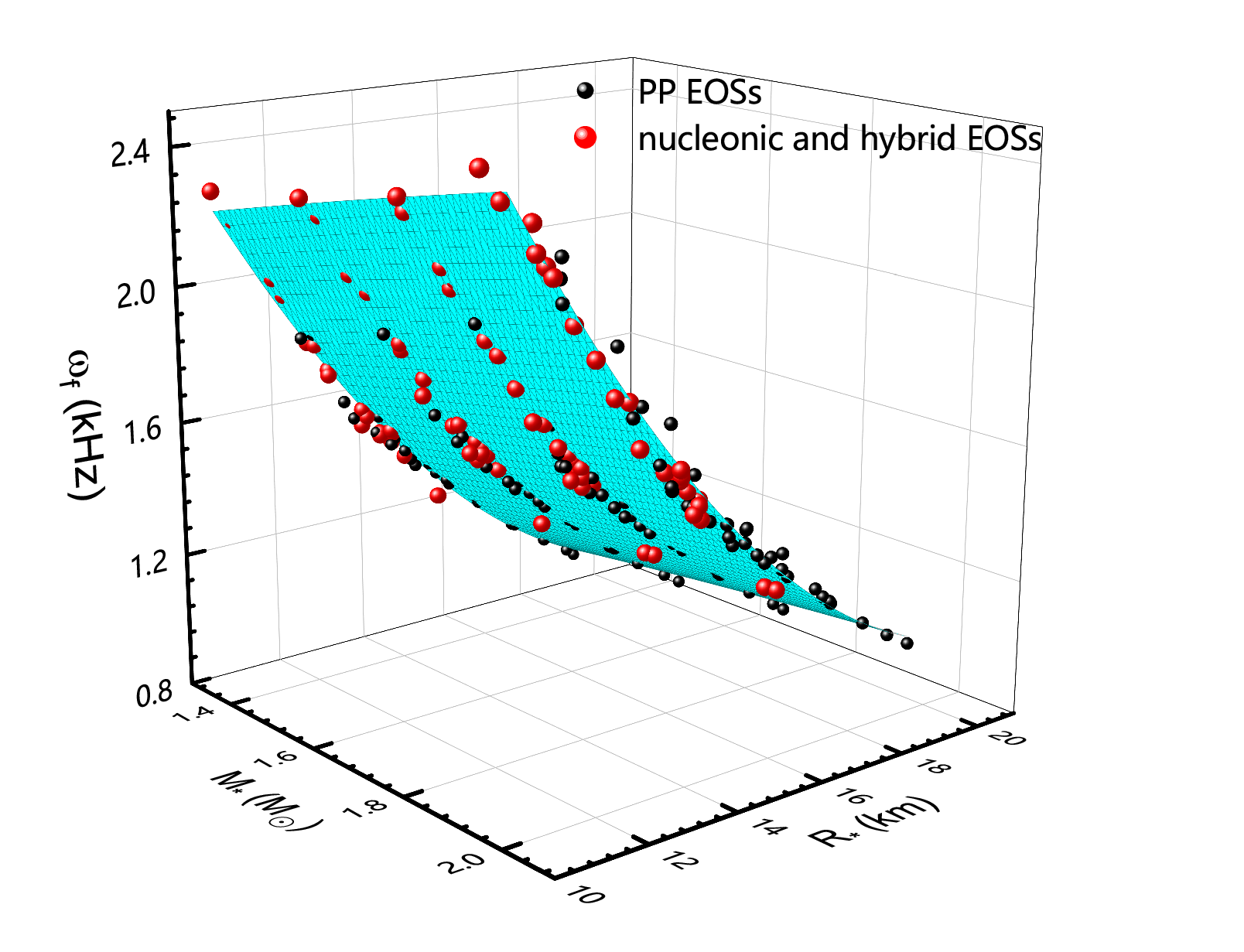}
    \caption{Similar to the Fig. \ref{2}, but for $\omega_f\ - \ \left( M_*,\ R_* \right)$.}
    \label{4}
\end{figure}

 Similar to the  treatment of universal relation $I_*\ -\ \left( M_*,\ R_* \right)$,   the frequencies $\omega_f$ of $f$-mode as a function of the stellar mass and radius are presented in  Figure \ref{4}.
  By fitting this surface, the  universal relation of  $\omega_f\ -\ \left( M_*,\ R_* \right)$ can be written as
\begin{equation}
\omega _f\left( \mathrm{kHz} \right)=4.66+0.838M_*-0.416R_*-0.0471{M_*}^2+0.0106{R_*}^2-0.0267M_*R_*, \label{3.3}
\end{equation}
where $M_*=\frac{M}{M_{\odot}}$ and $R_*=\frac{R}{\mathrm{km}}$.

\begin{figure}[H]
    \centering
    \includegraphics[scale=0.42]{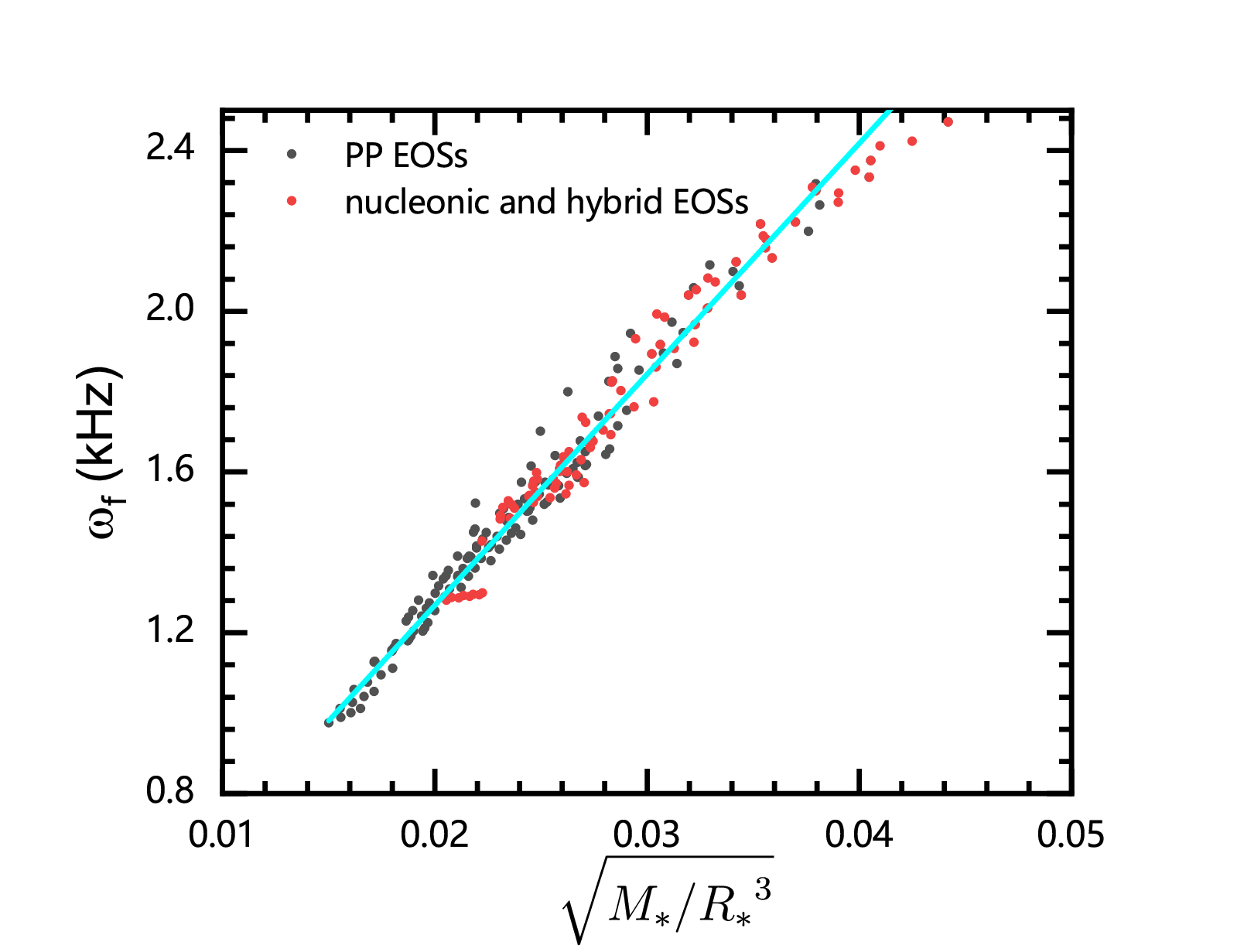}
    \caption{Similar to the Fig. \ref{3}, but for $\omega_f\ -\ \sqrt{M_*/{R_*}^3}$.}
    \label{5}
\end{figure}

Here we adopt the classical universal relation $\omega_f\ -\ \sqrt{M_*/{R_*}^3}$ as an example to compare the precision of the universal relations. Adopting the same set of EOSs, the universal relation between $\omega_f$ and $\sqrt{M_*/{R_*}^3}$ is displayed in Fig. \ref{5}. This universal relation can be fitted by
\begin{equation}
\omega _f\left( \mathrm{kHz} \right)=0.118+57.478\sqrt{M_*/{R_*}^3}. \label{3.4}
\end{equation}

The quantitatively comparison of the accuracy of these two universal relations shown in Table \ref{t2}. Again, the universal relation described by Eq. \ref{3.3} has a better precision.
 As a qualitative analysis, the lack of a separate $R_*$ term in Eq. \ref{3.4} may lead to the  higher relative error of $\omega_f\ -\ \sqrt{M_*/{R_*}^3}$ relation.

\begin{table}[H]
\caption{\textbf{Same as Table \ref{t1}, but for $\omega_f\ -\ \left( M_*,\ R_* \right)$ and $\omega_f\ -\ \sqrt{M_*/{R_*}^3}$}} \label{t2}
\centering
\begin{tabular}{lll}
\toprule
&~~~~~~~~~$\omega_f\ -\ \left( M_*,\ R_* \right)$&~~~~~~~~~$\omega_f\ -\ \sqrt{M_*/{R_*}^3}$ \\
\midrule
Mean $\left( \frac{1}{n}\sum_{i=1}^n{X_i} \right) $&~~~~~~~~~0.0698&~~~~~~~~~0.106 \\
\bottomrule
\end{tabular}
\end{table}

Similarly, based on Eq. (\ref{3.3}) and the astronomical observation data from NICER, the $f$-mode frequency of PSR J0030+0451 \cite{Miller_APJL2021} and PSR J0740+6620 \cite{Riley_APJL2021} can be constrained in  $\left[1.296-2.111\right]\ \textrm{kHz}$ and $\left[0.966-2.479\right]\ \textrm{kHz}$, respectively. The constraint of PSR J0030+0451 is basically consistent with the prediction ($1.67 - 2.18\ \textrm{kHz}$) of canonical neutron stars in Ref. \cite{Wen_PRC2019}. These constraints may provide useful guidance for the detection of gravitational radiation from the neutron star oscillation in the future.

\subsection{The universal relation for tidal deformation}\label{s2d}
\begin{figure}[H]
    \centering
    \includegraphics[scale=0.38]{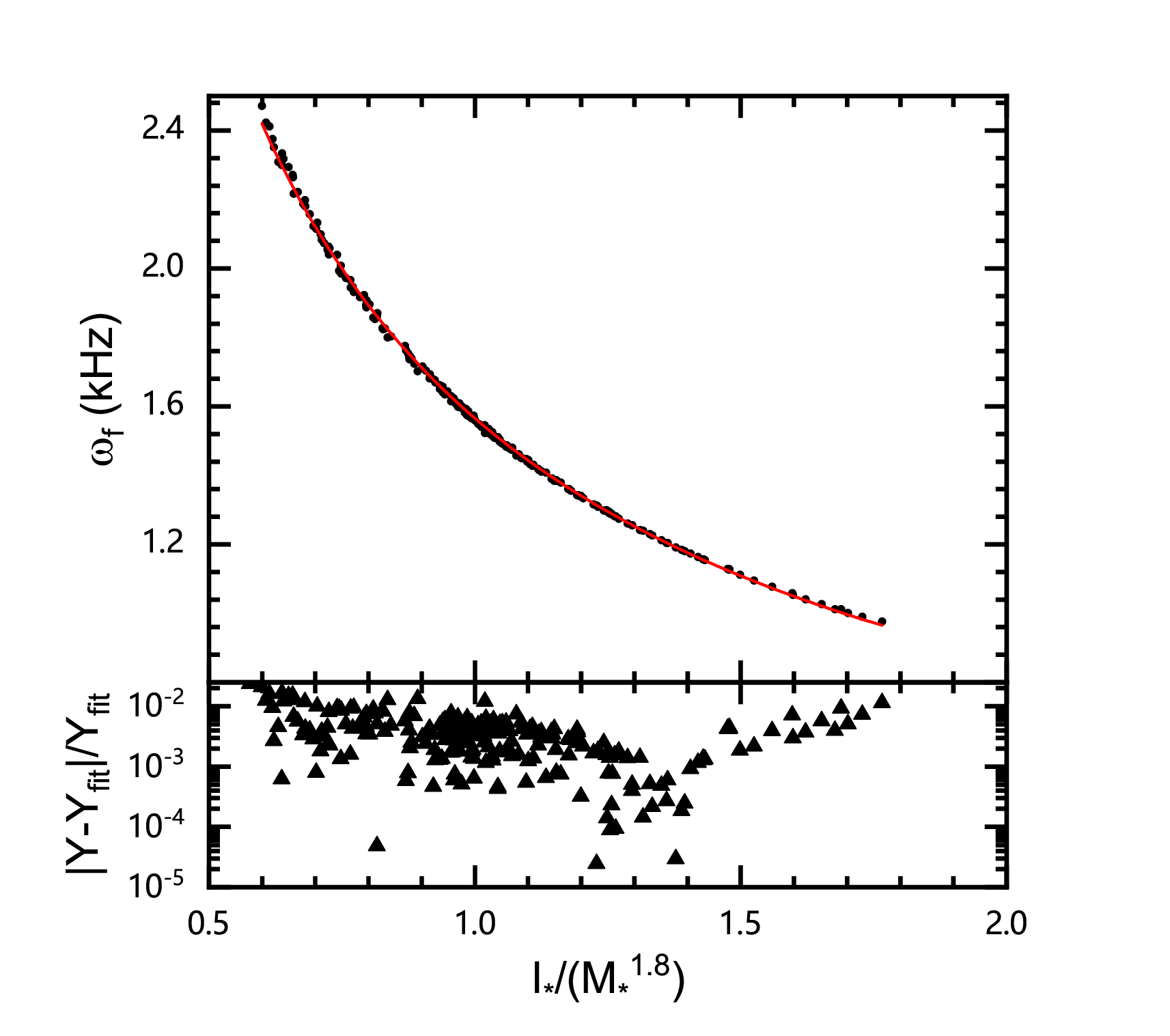}
    \caption{(Top) Universal relation between $I_*/{M_*}^{1.8}$ and $\omega_f$ for the same EOSs set together with the fitting curve.  (Bottom) Relative error between fitting curves and numerical results.}
    \label{6}
\end{figure}

Since the observation of GW170817, dimensionless tidal deformation $\varLambda$ has become an attractive observational property of neutron star contributing to constrain the EOS. Establishing the universal relation between $\varLambda$ and mass-radius provide a way to cross validate these three observable properties. Similar, through the mass-radius projection method, we can construct a precise universal relation between the tidal deformation and the mass-radius of neutron star.
Figure \ref{6} shows the universal relation between $\varLambda$ and the mass-radius. Here we express this universal relation by the following polynomial fitting
\begin{align}
\varLambda =108757.7751-299645.88054M_*+4654.63671R_*-5931.14445M_*R_*+289044.79877{M_*}^2 \nonumber \\
-80.98134{R_*}^2-127741.20618{M_*}^3+18.16444{R_*}^3+5789.65511{M_*}^2R_*-342.04328M_*{R_*}^2 \nonumber \\
+23162.31504{M_*}^4+0.35154{R_*}^4+295.79953{M_*}^2{R_*}^2-2493.99688{M_*}^3R_*-18.85851M_*{R_*}^3, \label{b1}
\end{align}
where $M_*=\frac{M}{M_{\odot}}$ and $R_*=\frac{R}{\mathrm{km}}$.

\begin{figure}[H]
    \centering
    \includegraphics[scale=0.45]{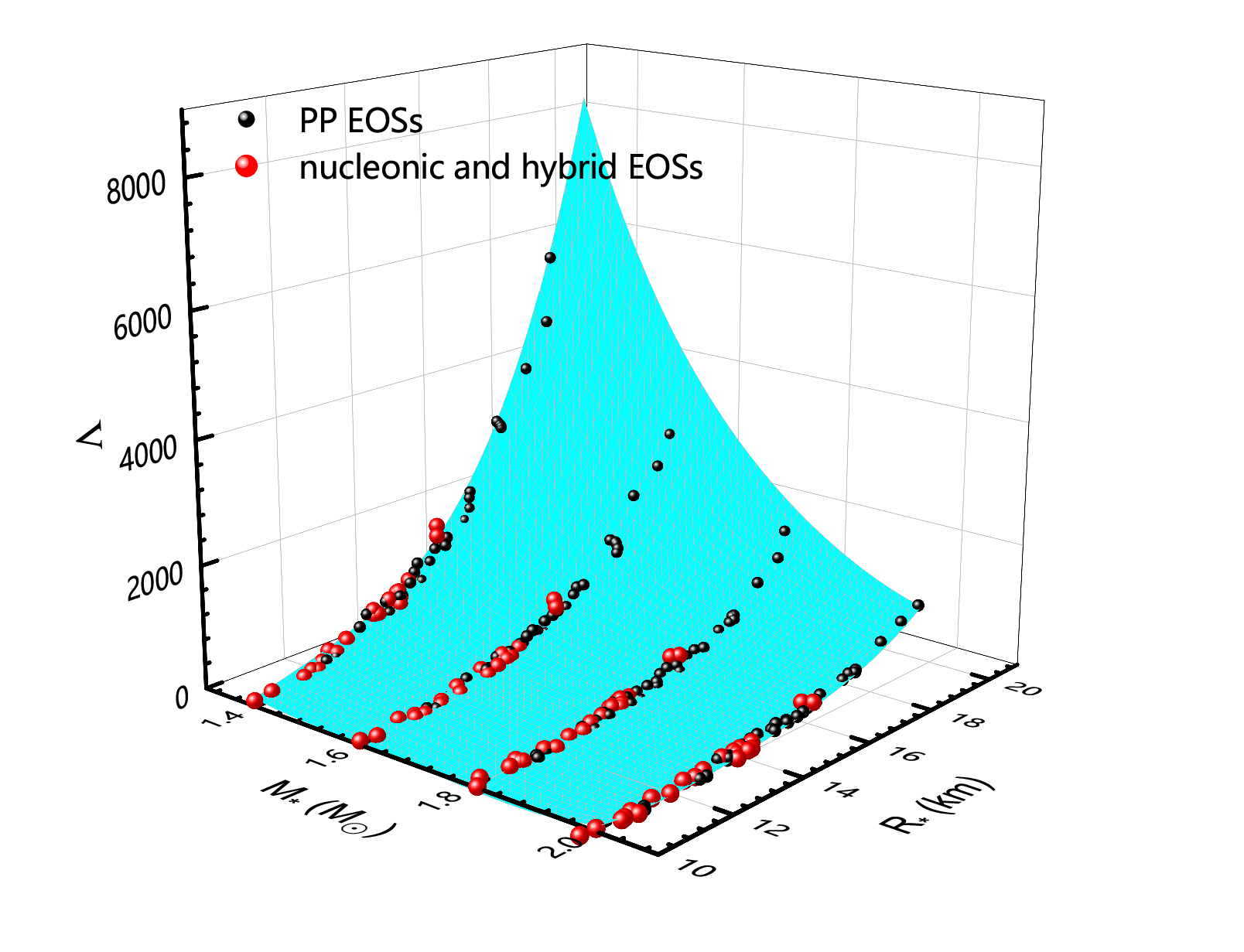}
    \caption{Similar to the Fig. \ref{2}, but for $\varLambda\ - \ \left( M_*,\ R_* \right)$.}
    \label{7}
\end{figure}

Here we also adopt the universal relation between the dimensionless tidal deformation $\varLambda$ and compactness $C$ as an example to compare the precision of the universal relations \cite{Maselli_PRD2013}. Adopting the same set of EOSs, the universal relation $\varLambda\ -\ C$ is displayed in Fig. \ref{7}. This universal relation can be fitted by
\begin{equation}
\varLambda =1.009\times 10^5-2.786\times 10^6C+2.878\times 10^7C^2-1.312\times 10^8C^3+2.223\times 10^8C^4. \label{b2}
\end{equation}

The quantitatively comparison of the accuracy of these two universal relations can be seen in Table \ref{t3}.  Similarly, the universal relation described by Eq. (\ref{b1}) has much better precision.

According to above fitting equation and the tidal deformation observational constraint $\varLambda _{1.4}=190_{-120}^{+390}$ of GW170817 \cite{Abbott_PRL2018},  the radius  of a canonical neutron star ($1.4\ M_{\odot}$) should be constrained in a range of $10.707_{-0.668}^{+1.967}\ \textrm{km}$. This result is in basic agreement with the prediction ($11\ -\ 12\ \mathrm{km}$) of Steiner \cite{Steiner_APJ2010}. In addition, by employing the mass and radius observational data from NICER, the tidal deformation for PSR J0030+0451 \cite{Miller_APJL2021} and PSR J0740+6620 \cite{Riley_APJL2021} can be constrained as $\left[ 284.713- 2833.076 \right]$ and $[<1885.358] $, respectively.

\begin{figure}[H]
    \centering
    \includegraphics[scale=0.42]{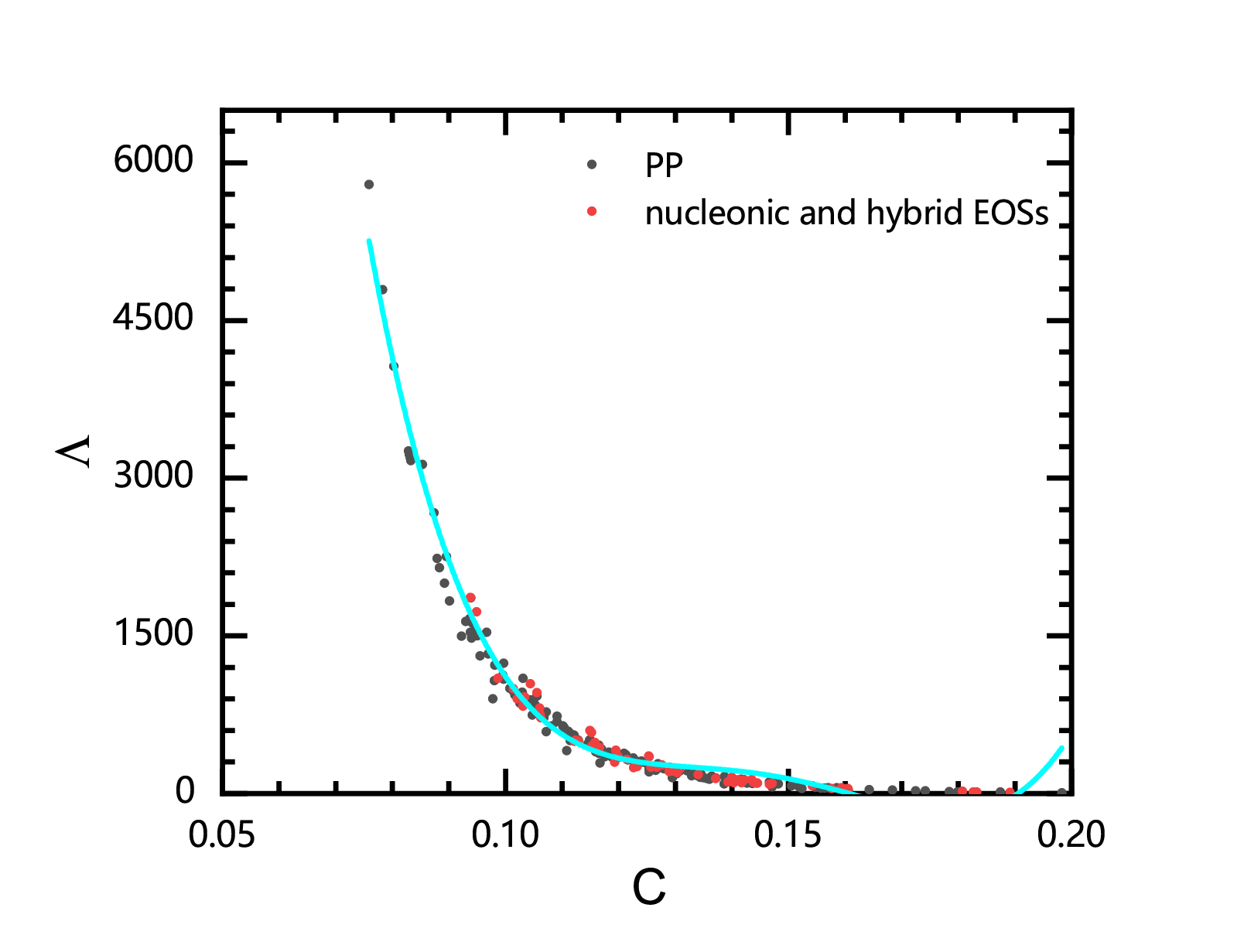}
    \caption{Similar to the Fig. \ref{3}, but for $\varLambda\ -\ C$.}
    \label{8}
\end{figure}

\begin{table}[H]
\caption{\textbf{Same as Table \ref{t1}, but for $\varLambda\ -\ \left( M_*,\ R_* \right)$ and $\varLambda\ -\ C$}} \label{t3}
\centering
\begin{tabular}{lll}
\toprule
&~~~~~~~~~$\varLambda\ -\ \left( M_*,\ R_* \right)$&~~~~~~~~~$\varLambda\ -\ C$ \\
\midrule
Mean $\left( \frac{1}{n}\sum_{i=1}^n{X_i} \right)$ &~~~~~~~~~0.0798&~~~~~~~~~0.213 \\
\bottomrule
\end{tabular}
\end{table}

\subsection{Constructing the universal relation between any two properties through mass-radius projection}\label{s2c}

Establishing the universal relation can be viewed as constructing a functional relation between two neutron star properties, i.e., $Y = F(X)$, where $X$ and $Y$ represent any properties of neutron star. Theoretically, if we have precise universal relations  $X\left( M_*,\ R_* \right)$ and $Y\left( M_*,\ R_* \right)$, then we can establish the universal relation between $X-Y$ by project one of the  3-dimensional surface to another surface. We called this projecting process as 3-dimensional affine transformation. Additionally, we provide the numerical solution in Appendix \ref{A}.

As an example, here we will construct the universal relation between the dimensionless moment of inertia $I_{*}$ and the  frequency $ \omega_f$ of $f$-mode by using the projection method, which based on   Eqs. ( \ref{3.1}) and (\ref{3.3}).
Operationally, to make the universal relation $\omega_f\left( \textrm{kHz} \right) = F(I_*)$ hold equals to make 3-dimensional surface $\omega_f \left( M_*,\ R_* \right)$ and surface $I_* \left( M_*,\ R_* \right)$ overlap. In mathematical transformation, the most likely transformation factors for this projecting operation should be ${M_*}^n$ and ${R_*}^n$. That is, we can roughly achieve the overlap of 3-dimensional surfaces $I_* \left( M_*,\ R_* \right)$ and $\omega_f \left( M_*,\ R_* \right)$ through Eq. (\ref{3.1}) (or Eq. \ref{3.3}) multiplied by the transformation factor.
Through analysis and trial, we find that if Eq. (\ref{3.1}) divided by the factor ${M_*}^{1.8}$, the projected surface $I_{*}(M_*,\ R_*)$  will overlap well with $\omega_f(M_*,\ R_*)$. Through this transformation, we can get a precise universal relation between  $I_{*}$ and $ \omega_f$ expressed as
\begin{equation}
\omega _f\left( \mathrm{kHz} \right)=1.565\left( I_*/{M_*}^{1.8} \right) ^{-0.851}. \label{3.5}
\end{equation}
This universal relation is illustrated in Figure \ref{8}. It is shown that the maximum relative error for all EOSs is within $2\ \%$ and the mean relative error is only $0.37\ \%$.

\section{Summary} \label{s3}

In this work, the mass-radius projection method is proposed to establish precise universal relations between neutron star global properties. Compared with the conventional universal relations, our newly constructed universal relations have higher precisions. As these new universal relations only are expressed by the mass and radius of neutron star,  the global properties (such as the moment of inertia, the frequency of $f$-mode, and the tidal deformation) can be directly constrained by the stellar mass and radius. Combing with the  mass and radius observation from NICER, these new universal relations provide an useful way  to precisely constrain the properties of the observed neutron stars. In addition, we offer a new perspective on constructing the universal relations, that is, a universal relation can be constructed through a 3-dimensional affine transformation based on the mass-radius projection. 

The established universal relations in this work is only applicable to the non-rotating static neutron stars. Theoretically, this projection method can also be extended to the rotating cases. For  rapid rotating neutron stars, as the rotating frequency $\varOmega$ will significantly effect the structure and properties \cite{Bejger_AA2013, Annala_PRX2022, Konstantinou_APJ2022},  the process of establishing the universal relation should be extended to a 4-dimensional projection, where the extra  dimensional variable is just the rotating frequency $\varOmega$.  Correspondingly, the transformation factors should be rewritten as ${M_*}^n$, ${R_*}^n$, and $\varOmega^n$. The establishment of the universal relation for the rotation condition through the use of $\varOmega$ factor can be found in Refs. \cite{Doneva_APJ2013, Chakrabarti_PRL2014, Kruger_PRD2023, Papigkiotis_PRD2023}.  We will further validate the above ideas in the future work.

\begin{acknowledgements}
This work is supported by NSFC (Grants No. 12375144, 11975101), Guangdong Natural Science Foundation (Grant No. 2022A1515011552).
\end{acknowledgements}

\appendix
\section{An alternative numerical solution for the new universal relation}\label{A}

In this section, we provide an approach to numerically solving the new universal relation. Establishing the universal relation is equivalent to finding the function $Y = F(X)$. In this context, we consider the universal relations $X\left( M_*,\ R_* \right)$ and $Y\left( M_*,\ R_* \right)$ as two-variable polynomials in the variables $M_*$ and $R_*$. Consequently, the process of establishing the universal relation can be viewed as analogous to performing multivariate polynomial division (MPD) \cite{Collins_JACM1971}. At this point, $Y = F(X)$ can be expressed as 
\begin{equation}
Y\left( M_*,\ R_* \right) =d\left( M_*,\ R_* \right) X\left( M_*,\ R_* \right) +r\left( M_*,\ R_* \right) ,
\end{equation}
where $d\left( M_*,\ R_* \right)$ represents the divisor and $r\left( M_*,\ R_* \right)$ is the remainder.
By performing MPD on Eq. (\ref{3.1}) and Eq. (\ref{3.3}), the universal relation between  $I_{*}$ and $ \omega_f$ can be expressed as
\begin{equation}
\omega_f\left( \mathrm{kHz} \right) = d\left( M_*,\ R_* \right) I_* + r\left( M_*,\ R_* \right),
\end{equation}
\begin{equation}
d\left( M_*,\ R_* \right) = -\frac{471}{4270},
\end{equation}
\begin{equation}
r\left( M_*,\ R_* \right) = \frac{327}{247000} M_* R_* + \frac{577111}{1067500} M_* + \frac{463}{38125} R_* - \frac{2017943}{4270000} R_* + \frac{3139801}{610000}.
\end{equation}
The mean relative error for this relation was 1.4\%. Due to the inherent errors in the universal relation of the mass-radius projection, the results obtained from MPD numerical solving may not be possible to achieve the lowest error.

$\\ \hspace*{\fill} \\$

\end{document}